\documentclass[structabstract]{aa}  
\usepackage{graphicx}
\usepackage{txfonts}	
\usepackage{units}
\usepackage{natbib} \bibpunct{(}{)}{,}{a}{}{,}
\usepackage{color}
\newcommand{\rt}{r_{\rm t}}
\newcommand{\rp}{r_{\rm p}}
\newcommand{\Mlst}{m_1}
\newcommand{\Msnd}{m_2}
\newcommand{\Mpwlw}{m_{\rm pw}}
\newcommand{\Mdust}{m_{\rm sr}}
\newcommand{\philst}{\phi_1}
\newcommand{\phisnd}{\phi_2}
\newcommand{\msec}{\rm ms^{-1}}
\newcommand{\mtar}{m_{\rm t}}
\newcommand{\mproj}{m_{\rm p}}
\newcommand{\phitar}{\phi_{\rm t}}
\newcommand{\phiproj}{\phi_{\rm p}}

\newcommand{\totder}[2]{\frac{{\rm d} #1}{{\rm d} #2}}
\newcommand{\partder}[2]{\frac{\partial #1}{\partial #2}}

\newcommand{\xa}{x_\alpha}					
\newcommand{\xb}{x_\beta}					
\newcommand{\va}{v_\alpha}					
\newcommand{\dab}{\delta_{\alpha\beta}}			

\newcommand{\stab}{\sigma_{\alpha\beta}}		

\newcommand{\rmd}{\mathrm{d}}				
\newcommand{\Sab}{S_{\alpha\beta}}			
\newcommand{\rhos}{\rho_{\rm s}}				
\newcommand{\phirbd}{\phi_{\rm RBD}}			
\newcommand{\pmean}{p_{\rm m}}				

\begin{document}
   \title{The four-population model: a new classification scheme for pre-planetesimal collisions}
   \titlerunning{The four-population model}

   \author{Ralf J. Geretshauser\inst{1}
          \and
          Farzana Meru\inst{1}
          \and
          Roland Speith\inst{2}
          \and
          Wilhelm Kley\inst{1}
          }

   \institute{Institut f\"ur Astronomie und Astrophysik, Abteilung Computational Physics,
   	    Eberhard Karls Universit\"at T\"ubingen, Auf der Morgenstelle 10, 72076
	    T\"ubingen, Germany\\
	     \email{ralf.j.geretshauser@uni-tuebingen.de}
	     \and
	    Physikalisches Institut, Eberhard Karls Universit\"at T\"ubingen, Auf der 	
   	    Morgenstelle 14, 72076 T\"ubingen, Germany
            }

   \date{}

 
  \abstract
   {Within the collision growth scenario for planetesimal formation, the growth step from centimetre-sized pre-planetesimals
   to kilometre-sized planetesimals is still unclear. The formation of larger objects from the highly porous pre-planetesimals may be halted by a combination of fragmentation in disruptive collisions and mutual rebound with compaction. However, the right amount of fragmentation is necessary to explain the observed dust features in late T Tauri discs. Therefore, detailed data on the outcome of pre-planetesimal collisions is required and has to be presented in a suitable and precise format.}
   {We wish to develop a new classification scheme broad enough to encompass all events with sticking, bouncing, and fragmentation contributions, accurate enough to capture the important collision outcome nuances, and at the same time simple enough to be implementable in global dust coagulation simulations. We furthermore wish to demonstrate the reliability of our numerical smoothed particle hydrodynamics (SPH) model and the applicability of our new collision outcome classification to previous results as well as our simulation results.}
   {We propose and apply a scheme based on the quantitative aspects of four fragment populations:  the largest and second largest fragment, a power-law population, and a sub-resolution population. For the simulations of pre-planetesimal collisions, we adopt the SPH numerical scheme with extensions for the simulation of porous solid bodies. On the basis of laboratory benchmark experiments, this model was previously calibrated and tested for the correct simulation of the compaction, bouncing, and fragmentation behaviour of macroscopic highly porous $\rm SiO_2$ dust aggregates.}
   {We demonstrate that previous attempts to map collision data were much too oriented on qualitatively categorising into sticking, bouncing, and fragmentation events. Intermediate categories are found in our simulations that are difficult to map to existing qualitative categorisations. We show that the four-population model encompasses all previous categorisations and in addition allows for transitions. This is because it is based on quantitative characteristic attributes of each population such as the mass, kinetic energy, and filling factor. In addition, the numerical porosity model successfully passes another benchmark test: the correct simulation of the entire list of collision outcome types yielded by laboratory experiments. As a demonstration of the applicability and the power of the four-population model, we utilise it to present the results of a study on the influence of collision velocity in head-on collisions of intermediate porosity aggregates. }
   {}


   \maketitle
%

\section{Introduction}

The primary way in which rocky planets are thought to form is by core accretion in protoplanetary discs consisting of gas and dust. Interaction between these components induces relative velocities and consequential mutual collisions among the initially micron-sized dust grains \citep[e.g.][]{Weidenschilling.1977, Weidenschilling.1993}. In this way, millimetre- to centimetre-sized fluffy dust aggregates grow by a simple hit-and-stick mechanism \citep[e.g.][]{Blum.2008}. The growth mechanism in the pre-planetesimal regime (centimetre to kilometre) is unclear and is addressed in this paper. Once a sufficient population of kilometre-sized objects has formed, gravitation-assisted accretion ensures the final assembly of full-sized planets \citep[e.g.][]{Goldreich.2004}. The planetary building blocks in this gravity-dominated regime are referred to as planetesimals. 

The growth step from pre-planetesimals to planetesimals is not well-constrained and the subject of extensive numerical and experimental effort. The difficulties to overcome in this step can be characterised with three types of barriers: fragmentation, drift, and bouncing barrier \citep[e.g.][]{Zsom.2010}. The most serious barrier is the fragmentation barrier: with increasing size, the relative velocities between pre-planetesimals increase and potentially lead to catastrophic disruption in mutual collisions. Often a velocity threshold of $\unit[1]{\msec}$ for disruptive events is assumed \citep{Blum.1993}. The threshold is mostly assumed to be as independent of other parameters such as porosity, mass ratio, and impact parameter \citep[e.g.][]{Guttler.2010}. With the assumption of this fragmentation threshold and power-law distributed fragments, \citet{Brauer.2008} find that dust coagulation is halted at centimetre or even millimetre sizes. \citet{Teiser.2009} proposed a growth model that is based mainly on the accretion of fragments $< \unit[1]{mm}$ by larger objects and therefore requires some amount of disruptive collisions. 

The second obstacle to planetesimal formation is the radial drift of larger pre-planetesimals \citep{Weidenschilling.1977b}. Owing to its pressure support, the gas in the disc rotates at a sub-Keplerian velocity, whereas the solid material lacks this pressure support and tends to rotate at a Keplerian velocity. As a consequence, the solid objects feel a headwind that causes an inward drift. For the minimum-mass solar nebula, it takes roughly a century for metre sized objects to drift  from \unit[1]{AU} into the star. In contrast, for centimetre- and kilometre-sized objects it takes $\unit[\sim 10^5]{yr}$. This is because small objects are strongly coupled to the motion of the gas, whereas large objects are almost decoupled. As a consequence, in the intermediate regime, metre-sized objects are quickly lost by accretion onto the host star or by photoevaporation in the hot zones close to it. Collective effects in the midplane of the protoplanetary disc might diminish the problem \citep[e.g.][]{Cuzzi.1993, Dominik.2007, Weidenschilling.2010}.

On the basis of empirical data showing the existence of rebound in dust collisions \citep{Blum.1993, Heisselmann.2007, Langkowski.2008, Weidling.2009, Guttler.2010}, \citet{Zsom.2010} introduced a possible new obstacle to growth: the bouncing barrier, where collision growth is halted at centimetre-sized pebbles. In the collisions, aggregates only become compacted in mutual collisions but do not grow any further. However, recent studies (Geretshauser et al., in prep., Wada et al., in prep.) indicate that the collision parameter space governed by bouncing may have been estimated to be too large by \citet{Guttler.2010} and that rebound is not as frequent as they inferred.

However, fragmentation is also necessary in discs. Observations \citep[e.g.][]{Natta.2007} suggest that sub-millimetre sized dust and millimetre- to centimetre-sized pebbles are present for as long as about $\unit[10^6]{yr}$ in T Tauri discs. However, \citet{Dullemond.2005} find that without fragmentation the disc is depleted of small grains within $\unit[10^3]{yr}$. This is commonly referred to as the grain retention problem. With coagulation simulations, \citet{Birnstiel.2009, Birnstiel.2010} tried to match the fragment distribution of disruptive events with the observational data.

To summarise, the formation of planetesimals requires the right amount of sticking, bouncing, and fragmentation to be consistent with observations. Therefore, collisions of pre-planetesimals have to be investigated as thoroughly as possible and their outcome has to be mapped as precisely as possible taking into account all the relevant parameters such as initial porosity, collision partner size, impact velocity, mass ratio, and rotation. To classify collision outcomes, \citet{Guttler.2010} compiled 19 experiments and mapped them to four types of sticking, two types of bouncing, and three types of fragmentation. However, owing to experimental restrictions they covered only small parts of the relevant parameter space. Many of their findings were not deduced from collisions between porous dust aggregates, but from dust collisions with a solid object. Collisions between aggregates larger than the decimetre size were not possible because of the limitations of their experimental apparatus. In addition, not all experiments could be carried out in protoplanetary disc conditions, i.e., in a vacuum and under microgravity.

An alternative to experiments is to study pre-planetesimal collisions numerically. To explore the region where experimental data is lacking, we perform simulations of pre-planetesimal collisions with our solid-body smoothed particle hydrodynamics (SPH) code, which was previously used by \citet{Schafer.2007} to emphasise the importance of precise material parameters and a thorough calibration of the implemented porosity model. \citet{Geretshauser.2010} expanded the porosity model and both calibrated and tested it extensively to simulate pre-planetary dust material based on laboratory benchmark experiments \citep{Guttler.2009}. The code and calibration procedure is briefly described in Sec.~\ref{sec:code-calibration}. In Sec.~\ref{sec:collision-types}, we show that the code can furthermore reproduce all sticking, bouncing, and fragmentation types proposed by \citet{Guttler.2010}. However, we find that their categorisation can also introduce unnecessary complexity and on some occasions may lack the required accuracy. It is possible that in a collision more than one process of sticking, bouncing, and fragmentation can take place and qualitative models do not make it clear whether the overall growth is positive, negative, or neutral. Therefore, we propose a new, simpler but at the same time more quantitative model for mapping collision data that is presented in Sec.~\ref{sec:new-model}. To show the applicability of this model, we present the first results of simulations of collisions between macroscopic objects consisting of realistic pre-planetesimal material in Sec.~\ref{sec:application}. We finally summarise our findings and discuss future work in Sec.~\ref{sec:discussion}.


\section{Code and calibration}
\label{sec:code-calibration}

For the simulations presented in this paper, we utilise the parallel SPH code \texttt{parasph} \citep{Hipp.2004, Schafer.2005, Schafer.2007}, which was calibrated to simulate porous dust material \citep{Geretshauser.2010}. This section briefly summarises the numerical model and calibration procedure.

\subsection{SPH and porosity model}

For collision simulations, SPH as a meshless Lagrangian particle method is superior to grid-based methods. This is because the continuous fluid or solid is divided into interacting mass packages, which serve as numerical sampling points and form a natural frame of reference for deformation and fragmentation (see reviews, e.g.\ by \citealt{Monaghan.2005} and \citealt{Rosswog.2009}). Originally developed for astrophysical fluid flows, SPH has also been extended to model solid material \citep[e.g.][]{Libersky.1991, Benz.1994, Randles.1996, Libersky.1997}.

Within the SPH scheme, the equations of continuum mechanics are solved in their Lagrangian form. In this framework, the continuity and momentum equations are given by
\begin{eqnarray}
\totder{\rho}{t} = - \rho \partder{\va}{\xa} \; , \\
\totder{\va}{t} = \frac{1}{\rho} \partder{\stab}{\xb} \; ,
\end{eqnarray}
where the Greek indices indicate the spatial components and the Einstein sum convention is applied. The quantities $\rho$ and $v$ are the density and velocity, respectively, and $\stab$ is the stress tensor which is defined as
\begin{equation}
\stab = - p \dab + \Sab \; ,
\end{equation}
where $p$ is the pressure, which accounts for pure hydrostatic compression or tension, and $\Sab$ is the deviatoric stress tensor, which represents pure shear. For the time evolution of the deviatoric stress tensor in the elastic regime, we adopt the approach of \citet{Benz.1994}, which uses the Jaumann rate form and follows Hooke's law. The elastic hydrostatic pressure is modelled as part of the porous equation of state (Eq.~\ref{eq:porosity-model}).

To model sub-resolution porosity, we relate the filling factor $\phi$ to the continuous quantities according to
\begin{equation}
\phi = \frac{\rho}{\rhos} \; ,
\end{equation}
where $\rhos = \unit[2000]{kg\,m^{-3}}$ is the density of the matrix material of $\rm SiO_2$ \citep[e.g.][]{Blum.2004}. Table~\ref{tab:material-parameters} shows the material parameters of the calibrated porosity model determined by \citet{Geretshauser.2010}. To model the plastic behaviour, we modified the approach by \citet{Sirono.2004}, who models plasticity with filling-factor-dependent strength quantities. The compressive strength $\Sigma(\phi)$, the tensile strength $T(\phi)$, and shear strength $Y(\phi)$ represent transition thresholds between the elastic and plastic regime. In particular, we adopt the relations (see \citealt{Guttler.2009} and \citealt{Geretshauser.2010})
\begin{equation}
\Sigma(\phi) = \pmean \left( \frac{\phi_{\rm max} - \phi_{\rm min}}{\phi_{\rm max} - \phi} - 1 \right)^{\Delta\,\ln{10}} \; ,
\label{eq:compressive-strength}
\end{equation}
for $\phi_{\rm min} + \varepsilon < \phi < \phi_{\rm max}$ and $\varepsilon = 0.005$. The quantities $\phi_{\rm min} = 0.12$ and $\phi_{\rm max} = 0.58$ denote the minimum and maximum filling factors, respectively, in the compressive strength relation. However, the material can exceed both of these values. The power of the compressive strength relation is $\ln(10)$ times the parameter $\Delta$ with $\Delta = 0.58$ \citep[see][for a study of this parameter]{Geretshauser.2010}, which can be found together with the other material parameters in Table \ref{tab:material-parameters}. For $\phi \le \phi_{\rm min} + \varepsilon$, the compressive strength relation is continuously extended by the constant function $\Sigma(\phi) = \Sigma(\phi_{\rm min} + \varepsilon)$ and for $\phi_{\rm max} \le \phi$ we set $\Sigma(\phi) = \infty$. The tensile strength is given by
\begin{equation}
T(\phi) = \unit[- 10^{2.8 + 1.48\phi}]{Pa} \; ,
\end{equation}
and the shear strength is the geometric mean of the compressive and tensile strength
\begin{equation}
Y(\phi) = \sqrt{\Sigma(\phi) | T(\phi) |} \; .
\end{equation}
Hence, the full equation of state for the hydrostatic pressure reads
\begin{equation}
p (\phi) = \left\{
\begin{array}{ll}
\Sigma(\phi) & \mathrm{for}~\phi_c^+ < \phi \\
K(\phi_0') & \mathrm{for}~\phi_c^- \le \phi \le \phi_c^+\\
T(\phi) & \mathrm{for}~\phi < \phi_c^-
\end{array} \right. \; ,
\label{eq:porosity-model}
\end{equation}
where $\phi_c^+ > \phi_c^-$, and $\phi_c^+$ and $\phi_c^-$ are critical filling factors. The value of $\phi_c^+$ marks the transition between elastic and plastic compression and $\phi_c^-$ defines the transition between elastic and plastic tension. The filling factors in-between the critical values represent the elastic regime. The quantity $\phi_0'$ is the reference density at vanishing external pressure and $K(\phi)$ is the filling-factor-dependent bulk modulus given by
\begin{equation}
K (\phi) = K_0 \left( \frac{\phi}{\phirbd} \right)^\gamma \; ,
\end{equation}
where $\gamma = 4$, $K_0$ is the bulk modulus of an uncompressed random ballistic deposition (RBD) dust sample with $\phirbd = 0.15$ \citep[e.g.][]{Blum.2004} such that $K_0 = K(\phirbd) = \unit[4.5]{kPa}$, and the shear modulus is given by $\mu (\phi) = 0.5 K(\phi)$.

Plasticity is modelled such that once the filling factor exceeds the critical values $\phi_c^+$ or $\phi_c^-$, the hydrostatic pressure, which is computed according to the elastic equation of state (second line in Eq.~\ref{eq:porosity-model}), is reduced to $\Sigma(\phi)$ or $T(\phi)$, respectively. For the reduction in pure shear, the deviatoric stress tensor $\Sab$ is limited according to the van Mises plasticity given by \citep[e.g.][]{Benz.1994}
\begin{equation}
\Sab \rightarrow f \Sab \; ,
\end{equation}
where $f = \min \left[ Y^2 / 3 J_2, 1 \right]$, $J_2 = \Sab \Sab$ is the second irreducible invariant of the deviatoric stress tensor, and $Y = Y(\phi)$ is the shear strength.

\subsection{Calibration procedure}

The above porosity model was carefully calibrated and tested in \citet{Geretshauser.2010} with the aid of three benchmark experiments \citep{Guttler.2009}: (1) the compaction of dust by a glass sphere, (2) the rebound of a dust aggregate from a solid surface, and (3) the fragmentation of a dust aggregate colliding with a solid surface. This is the most elaborate material calibration which is available for astrophysically relevant macroscopic dust aggregates. All benchmark experiments were carried out in the laboratory as well as in simulations.

Because compaction of porous material is a very efficient energy dissipation mechanism that allows dust aggregates to stick rather than break apart, in our first experiment we tested the compression properties of the model, ensuring that both the dynamic compressive strength and the shear strength relations were calibrated.

Since rebound is discussed as an obstacle to planetesimal formation, the second benchmark experiment tested the bouncing properties of a dust aggregate. With the aid of this benchmark setup, we calibrated the bulk modulus as a characteristic quantity of the elastic properties. We also confirmed the relation of the compressive strength which is a characteristic quantity for the energy dissipation during compaction.

For pre-planetesimal collisions that end in fragmentation, the largest fragment and the mass distribution of the fragments is highly important for subsequent collisions. For this reason, our last calibration experiment tested the correct reproduction of the fragmentation properties of the porosity model. The bulk modulus, which is the quantity that has the strongest influence on the fragment mass distribution, was also confirmed.

After the successful calibration procedure, we are able to correctly reproduce the compaction, bouncing, and fragmentation behaviour of porous $\rm SiO_2$ dust aggregates quantitatively with one consistent set of material parameters as shown in Table \ref{tab:material-parameters}. These material parameters are used for all simulations discussed in the following.

\begin{table}
\centering
\begin{tabular*}{\hsize}{l@{\extracolsep{\fill}}c c l}
\hline\hline
Physical Quantity & Symbol & Value & Unit \\
\hline
\\
Bulk modulus		& $K_0$		& 4,500		& Pa \\
Bulk modulus exponent & $\gamma$ & 4 & - \\
Matrix density		& $\rhos$	& 2,000			& $\rm kg\,m^{-3}$ \\
RBD filling factor	& $\phirbd$	& 0.15			& - \\
$\Sigma$ mean pressure	& $\pmean$		& 260			& Pa \\
$\Sigma$ max.\ filling factor& $\phi_{\rm max}$	& 0.58			& - \\
$\Sigma$ min.\ filling factor& $\phi_{\rm min}$	& 0.12			& - \\
$\Sigma$ power parameter	& $\Delta$	& 0.58			& - \\
\hline
\end{tabular*}
\caption{$\rm SiO_2$ dust material parameters used in the porosity model. RBD stands for the random ballistic deposition dust sample by \citet{Blum.2004} and $\Sigma = \Sigma (\phi)$ is the compressive strength.}
\label{tab:material-parameters}
\end{table}


\section{Reproducing sticking, bouncing, and fragmentation collision types}
\label{sec:collision-types}

\begin{figure*}
   \resizebox{\hsize}{!}
            {\includegraphics{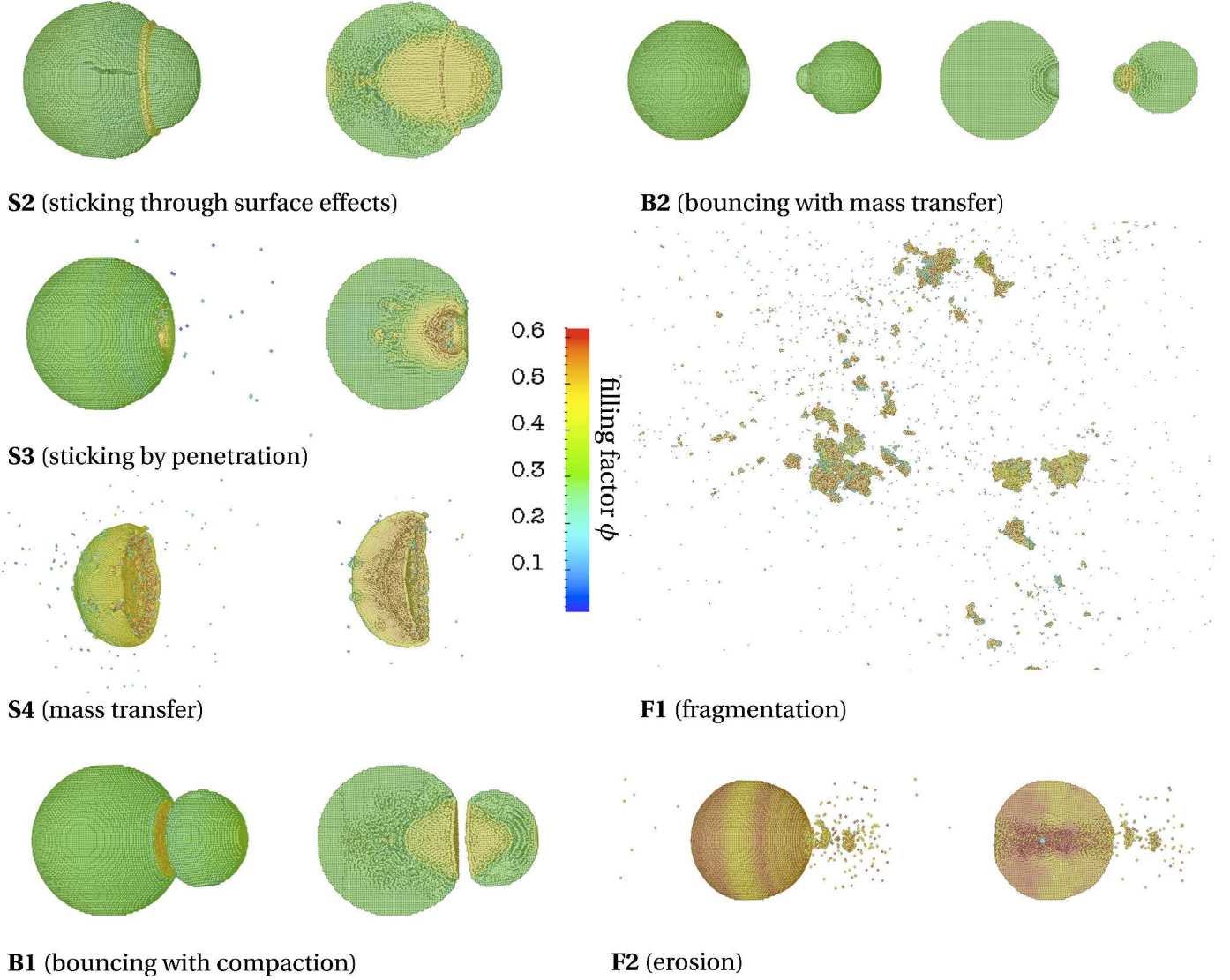}}
      \caption{The outcome of our simulations of pre-planetesimal collisions encompasses all sticking (S), bouncing (B), and fragmentation (F) types proposed by \citet{Guttler.2010}. The initial configuration for each simulation is a sphere with radius $\rt = \unit[10]{cm}$ as resting target and a sphere with $\rp = \unit[6]{cm}$ as projectile except for S3 and F2, where $\rp = \unit[2]{cm}$. The colour code indicates the filling factor $\phi$. Both objects are initially set up with $\phi = 0.35$ except for the F2 case, where $\phi = 0.55$. The simulations are carried out with different impact velocities and the snap shots are taken at different times. The details are given in the text.}
         \label{fig:collision-types}
\end{figure*}

\citet{Guttler.2010} compiled a large number of experiments with dust aggregates and suggested categorising them into four types of sticking, two types of bouncing, and three types of fragmentation. The sticking events were subdivided according to sticking mechanisms (hit-and-stick for micron sized monomers, surface effects, or deep penetration) and a quantitative criterion (mass transfer). Growth neutral bouncing with compaction and bouncing with mass transfer were distinguished in terms of rebound. Fragmentation was classified according to its degree (complete fragmentation or erosion) and whether some sticking was involved (fragmentation with mass transfer).

In this section we show that our code and its underlying calibrated porosity model can reproduce these empirical outcome types in numerical simulations. Images of the final result of each example simulation are presented in Fig.~\ref{fig:collision-types} according to this categorisation. The end times of the simulations are chosen by visual control of the fragment evolution and particle velocities close to the thermal equilibrium for each fragment. For each category, an exterior view (left) and a cut through the centre of the resulting aggregate(s) (right) is shown (except for type F1). The target is always a dust aggregate with radius $\rt = \unit[10]{cm}$ modelled using 238,238 SPH particles. The projectiles are modelled using 1,905 and 51,477 SPH particles for projectile radii $\rp = \unit[2]{cm}$ and $\rp = \unit[6]{cm}$, respectively, and placed on a cubic lattice with edge length \unit[2.6]{mm}. The collision partners are aligned head on with a shift of half a lattice constant into each direction perpendicular to the collision axis to avoid particle interpenetration. In all simulations (except F2), both objects are set up with $\phi=0.35$ and the masses of the target, and the \unit[6]{cm}-, and \unit[2]{cm}-projectiles are \unit[2.93]{kg}, \unit[0.63]{kg}, and \unit[0.023]{kg}, respectively. An overview of the initial conditions of the simulations is given in Table \ref{tab:collision-type-table}. The colour in Fig.~\ref{fig:collision-types} indicates the filling factor $\phi$. We now discuss the simulation outcomes in detail. We adopt the notation by \citet{Guttler.2010}.

\emph{S1 (hit \& stick)} is not represented in Fig.~\ref{fig:collision-types} as it is not applicable within the continuous SPH scheme. This category describes the outcome of collisions in which microscopic fractal dust aggregates are involved. In this regime, the continuum limit, which is a fundamental assumption of our model, is invalid. We focus on collisions of macroscopic dust since for the fractal regime the numerical and empirical basis is profound.

\emph{S2 (sticking through surface effects)} The example shows the result of the impact of an aggregate with radius $\rp = \unit[6]{cm}$ and collision velocity $v_0 = \unit[2.0]{\msec}$. The image shows the situation \unit[250]{ms} after the impact. The filling factor of the outer shell of the aggregates remains nearly unchanged as indicated by the exterior view. In contrast, large parts of the interior are compressed to $\phi \sim 0.45$. The target and the projectile merge to form one object. The initial setup matches that used for the B2 case but has a higher impact velocity. As a consequence, the contact area is larger and compressed to higher filling factors. This leads to a higher tensile strength in this region, which prevents the objects from rebounding from each other. This situation is referred to as sticking through surface effects. A single aggregate remains and no fragments are ejected.

\emph{S3 (sticking by deep penetration)} is found in our simulations if the projectile is sufficiently smaller than the target. In the example case, the projectile has a radius of $\rp = \unit[2]{cm}$ and hits the target with $v_0 = \unit[10.0]{\msec}$. The image is taken \unit[388]{ms} after the impact. A small crater is formed on the target and a small number of single SPH particles are ejected. This characterises deep penetration. The crater is visible in the cross-section. Material is compressed to a maximum of $\phi \sim 0.52$ down to a depth of about one target radius.

\emph{S4 (mass transfer)} occurs if the projectile is large and moves fast enough to stick but still slow enough to not fully disrupt the target. As an example, we choose an impact of a \unit[6]{cm} aggregate onto a target with the same filling factor. The impact velocity is $v_0 = \unit[10.0]{\msec}$ and the image shows the result after \unit[50]{ms}. During the collision, the target is highly deformed and compressed to filling factors of $\phi \sim 0.45$ to $0.52$. Single SPH particles are ejected. Most of the projectile merges, which accounts for the mass transfer. The final mass of the largest fragment is \unit[3.56]{kg}.

\emph{B1 (bouncing with compaction)} is only seen in the case of objects of medium and low porosity at low collision speeds. Highly porous objects have an extremely low compressive strength. As a consequence, low pressures suffice to plastically deform the target and the highly porous aggregates do not gain enough elastic loading for bouncing. In our example, the projectile is \unit[6]{cm} in radius. It impacts at $v_0 = \unit[1.0]{\msec}$. In Fig.~\ref{fig:collision-types}, the final state at \unit[500]{ms} is shown. In the collision, both objects are flattened at the impact site and the elastic loading is sufficient to make them rebound after the impact. Part of the interior of the target and projectile is compressed to $\phi \sim 0.45$. We note that this is exactly the same setup as in the S2 case but with a lower impact velocity. The bouncing event is a result of a smaller contact area and a lower tensile strength in this region, which is indicated by the lower filling factor. As a consequence, the two objects rebound. A close investigation reveals that in nearly every simulation with bouncing, a small amount of material is transferred between the projectile and target. Thus, it is very unlikely that pure growth neutral bouncing without any mass transfer exists. In this particular simulation, material of $\sim \unit[0.54]{g}$ is transferred from the projectile to the target. Hence, the result could also be classified into the next category.

\emph{B2 (bouncing with mass transfer)} occurs in simulations where the compressive strength is sufficiently large to allow for elastic loading and consequential bouncing, and where the tensile strength is small enough for the impactor to rip out a small fraction of mass from the target. This happens at very low impact speeds. The final state at \unit[1.7]{s} after the collision is the result of a collision with $v_0 = \unit[0.2]{\msec}$. The projectile has $\rp = \unit[6]{cm}$. During the impact, a small region of target and projectile is compressed to $\phi \sim 0.45$. In the consequential rebound, this region sticks to the projectile and is ripped out of the target. The remaining crater on the target can be seen in the cross-section. In this collision, \unit[29.5]{g} are transferred from the target to the projectile.

\emph{F1 (fragmentation)} is generally the outcome of collisions with high impact velocities. Highly porous objects effectively dissipate energy by deformation because of their low compressive strength but they are also easy to disrupt because of their low tensile strength. In contrast, objects of low porosity have high tensile strengths, but lack the ability to dissipate large amounts of kinetic energy. The degree of fragmentation is therefore strongly porosity dependent. Our example shows the result of a collision between objects, where the projectile features $\rp = \unit[6]{cm}$. The impact speed is $v_0 = \unit[17.5]{\msec}$. During the collision, both the target and the projectile completely shatter and the result of this collision (shown only in exterior view \unit[800]{ms} after the impact) is a continuous fragment distribution, whose masses range from \unit[285]{g}
down to single SPH particles of \unit[12]{mg}. The fragments consist of target and projectile material that are indistinguishably combined together. The filling factors of the fragment distribution are $\phi \sim 0.54$, which is close to the maximum filling factor of 0.58 (see Table \ref{tab:material-parameters}).

\emph{F2 (erosion)} is observed particularly for high filling factors and small projectile radii. The initial setup for the example case involves two objects with $\phi = 0.55$. The target and projectile masses were \unit[4.60]{kg} and \unit[0.037]{kg}, respectively. The projectile radius is $\rp = \unit[2]{cm}$ and the impact velocity $v_0 = \unit[20.0]{\msec}$. During the intrusion of the projectile, fragments consisting of a small number of SPH particles are ejected opposite the impact site. The figure shows the situation $\unit[438]{ms}$ after the collision, where small fragments and SPH particles are ejected from the crater. In this collision, \unit[63.2]{g} of material are eroded from the target. The projectile intrudes to the centre of the target and the intrusion channel is clearly visible in the cross-section. Because of the high compressive strength elastic deformation also takes place on the target. The figure shows a snap shot of a decompression wave of lower filling factor, which appears as a vertical ring on the target.

The category \emph{F3 (fragmentation with mass transfer)} contains conceptual difficulties. The first reason is that in the simulation outcomes of the F1 and F2 categories, some mass is always transferred to the target. Secondly, the demarcation between sticking with mass transfer (S4) and F3 is unclear and a continuous transition between these categories is expected. This problem can be solved by our new model proposed below. Our simulations incorporate category F3 by reproducing the S4 and F1 types. 

In conclusion, our code is not only capable of quantitatively reproducing sticking, bouncing, and fragmentation in general \citep{Geretshauser.2010} but also of correctly simulating the sub-types for each macroscopic collision outcome. However, we also experienced difficulties in classifying the results of our simulations according to the model of \citet{Guttler.2010}. This motivates our wish to find a new approach.

\begin{table}
\centering
\begin{tabular*}{\hsize}{l @{\extracolsep{\fill}} c c c c c c}
\hline\hline
Type & $v_0$ [$\msec$] & $\rt$ [cm] & $\mtar$ [kg] &$\rp$ [cm] & $\mproj$ [kg] & $\phi_{\rm t/p}$  \\
\hline
\\
S2	& 2.0 & 10 & 2.93 & 6 & 0.63 & 0.35 \\
S3	& 10.0 & 10 & 2.93 & 2 & 0.023 & 0.35 \\
S4	& 10.0 & 10 & 2.93 & 6 & 0.63 & 0.35 \\
B1 	& 1.0 & 10 & 2.93 & 6 & 0.63 & 0.35 \\
B2	& 0.2 & 10 & 2.93 & 6 & 0.63 & 0.35 \\
F1	& 17.5 & 10 & 2.93 & 6 & 0.63 & 0.35 \\
F2	& 20.0 & 10 & 4.60 & 2 & 0.037 & 0.55 \\
\hline
\end{tabular*}
\caption{Initial parameters of the simulations shown in Fig.~\ref{fig:collision-types}. The quantities are: collision velocity $v_0$, target radius $\rt$ and mass $\mtar$, projectile radius $\rp$ and mass $\mproj$, and the filling factor of the collision partners $\phi_{\rm t/p}$.}
\label{tab:collision-type-table}
\end{table}


\section{A new model for mapping collision outcomes}
\label{sec:new-model}

We introduce a new model to classify the outcome of pre-planetesimal collisions. We first describe its structure. We then show that all sticking, bouncing, and fragmentation events of the previous section can be incorporated into the new model.

\subsection{The motivation behind a new classification scheme}

The model of \citet{Guttler.2010} was developed based on particular laboratory setups. In several of these, one of the collision partners was not a dust aggregate but a solid surface or a glass bead, which itself cannot fragment. Therefore, applying this categorisation to pre-planetesimal dust collisions, which here are solely carried out as simulations between dust aggregates, leads to some difficulties.

In our simulations, the collisions of S3 (sticking by penetration) always produce ejected dust. Therefore, it is unclear whether they should be classified as S3 or rather S4 (mass transfer). On the other hand, S2 (sticking through surface effects) involves also a mass transfer but without the production of fragments. It can also be expected that there is a continuous transition from S2 to S3 with increasing impact velocity. The demarcation between the sticking types is also conceptually difficult to make. The types S1 to S3 are distinguished on the basis of the sticking mechanism, whereas for S4 the criterion is an increase in target mass and the generation of some fragments. In the context of pre-planetesimal growth, the exact sticking mechanism is of minor importance. The distinction between growing and disruptive events is simply given by comparing the \emph{largest fragment} before and after the collision.

Applying the bouncing categories to collision data also causes some difficulties. Pure bouncing with compaction (B1) events are never seen in our simulations. Analysing the final mass of both collision partners after the impact reveals that some mass is always transferred either from the projectile to the target or vice versa. In the first case, growth occurs and the result should be categorised under S4 with a growing target, but with one remaining fragment instead of a fragment distribution. If mass is transferred to the projectile, then the event belongs to B2 (bouncing with mass transfer). However, since the largest object is losing mass, B2 is also a type of erosion (F2), where instead of a fragment distribution only one large fragment (the enlarged projectile) is present. Since bouncing may be an obstacle to planetesimal growth \citep{Zsom.2010}, the effect of bouncing has to be included in any collision map. However, instead of distinguishing between two types of bouncing, we find that it is sufficient to distinguish between the mass of the \emph{largest} and the \emph{second largest fragment}.

Analysing the numerical results for disruptive events, it is also hard to distinguish between types F1 (fragmentation) and F3 (fragmentation with mass transfer). As already mentioned above, in any fragmenting collision, mass is transferred from the projectile to the target and the fragments consist of both projectile and target material. In eroding events (F2), mass is also transferred to the target and a fragment distribution is produced. Once again the continuous transition between the three fragmentation types and between sticking and fragmentation events cannot adequately be described by the given model. In our view, it is sufficient to characterise the outcome of disruptive events in terms of the size of the largest fragment (which for F2 is much larger than the other fragments) and a power-law distribution of the remaining fragments. Thus, for correct mapping of any combination of sticking, bouncing, and fragmentation, the following objects have to be considered: both a \emph{largest} and a \emph{second largest fragment} as well as a \emph{power-law fragment distribution}.

This discussion shows that the classification of different sticking, bouncing, and fragmentation events is not sufficient. This is because many intermediate events exist. For example, a bouncing event can involve some sticking and fragmentation, whereas a sticking event can also involve some fragmentation. The model by \citet{Guttler.2010} tries to capture these intermediate events. However, it is based on physical mechanisms rather than collision outcome. Owing to its discrete approach, the model generates some unnecessary complexity and cannot easily describe transitions between categories. To improve on this, we propose a model that is based solely on quantitative aspects. It distinguishes between four types of fragment populations characterised by continuous quantities such as their mass, filling factors, velocities, and size. In this way, we are able to map all types of sticking, bouncing, and fragmentation events but also model continuous transitions between these types. 

\subsection{Fragment populations}
\label{sec:frag_pop}

\begin{figure*}
   \resizebox{\hsize}{!}
            {\includegraphics{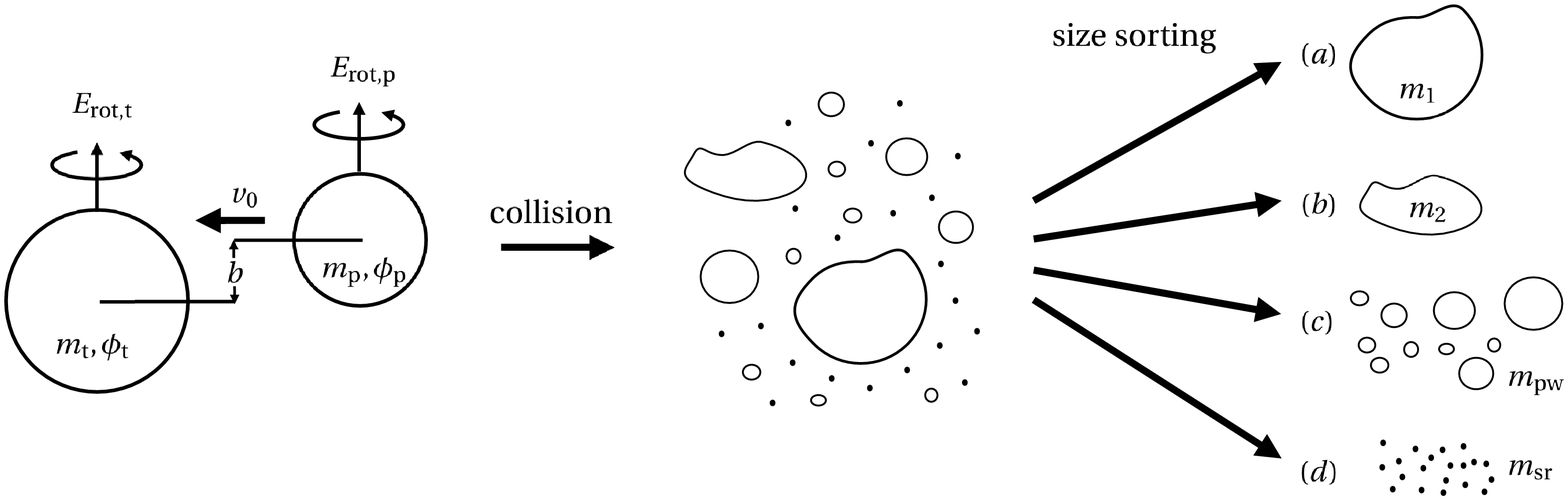}}
      \caption{Illustration of the four-population model. The left side displays the situation before the collision: two pre-planetesimals collide with impact velocity $v_0$ and impact parameter $b$. The target and projectile are characterised by their defining quantities, e.g., mass $m$, filling factor $\phi$, and rotational energy $E_{\rm rot}$. Depending on these parameters, an outcome population is generated in the collision (middle). The four-population model distinguishes the fragments according to their mass. The population classes are: (a) the largest fragment, (b) the second largest fragment, (c) the power-law population, and (d) the sub-resolution population. This categorisation is sufficient to describe the outcome types of Fig.~\ref{fig:collision-types} as well as intermediate outcomes.}
         \label{fig:collision-model}
\end{figure*}

The fundamental idea of the new classification is that any outcome of a collision can adequately be modelled by distinguishing four kinds of fragment ``populations''. This is illustrated in Fig.~\ref{fig:collision-model}. Example input parameters for the collision include the masses of the target (label ``t'') $\mtar$ and projectile (label ``p'') $\mproj$, their filling factors $\phitar$ and $\phiproj$, their rotational energies $E_{\rm rot, t}$ and $E_{\rm rot, p}$, their impact parameter $b$ and, most importantly, their collision velocity $v_0$ (see Fig.~\ref{fig:collision-model}, left). Depending on these parameters, the collision between two bodies produces a well-defined outcome of populations. To be able to map sticking, bouncing, and fragmentation events, it is suitable to sort this distribution according to the mass of the fragments in the respective population. For these, output parameters such as mass, filling factor, rotational energy, and velocity are determined as exact or averaged values, or as distributions depending on the population. Specifically, the new scheme consists of the following four types: (1) largest fragment, (2) second largest fragment, (3) power-law population, and (4) sub-resolution population.

\emph{(a) Largest fragment:} Comparing the mass of the largest fragment before and after a collision distinguishes between positive and negative growth. The characteristic output quantities are supplied. 

\emph{(b) Second largest fragment:} This enables bouncing events to be mapped. In pure bouncing events, the largest and the second largest fragment will be the only members of the fragment population. The quantities describing this fragment are supplied. The second largest fragment only exists if it consists of more than a single SPH particle. 

\emph{(c) Power-law population:} in disruptive collision events in particular, the mass distribution of the fragments can be modelled by a power-law to describe the fragment distribution. Instead of describing each fragment of the power-law population with quantities as done previously for the largest and second largest fragment, the number of parameters is reduced significantly by utilising distribution functions.

\emph{(d) Sub-resolution population:} This population is introduced for numerical reasons. Fragments of the sub-resolution population consist only of a single SPH particle. The existence of this population represents the resolution limit of the simulations and therefore represents only an upper limit to the smallest dust fragments, which are produced in a collision. This population is also convenient to control sufficient resolution. The SPH numerical scheme is capable of simulating objects of metre size and more. However, because of limited computational resources, single SPH particles then represent objects of centimetre size and larger. The sub-resolution population keeps track of these objects. It is convenient to describe this population with averaged characteristic quantities. We note that also in laboratory experiments a sub-resolution population exists, for which the fragment distribution is determined by high speed cameras.

With this approach, we considerably reduce the complexity of the sticking, bouncing, and fragmentation classification from nine types of collision outcomes to four fragment populations. Moreover, our model is based solely on continuous quantities characterising the collision outcome and not on physical mechanisms. This allows us to model any mixed types of collisions and transitions between growth and disruption with the necessary accuracy.

Compared to fragmentation data that are mapped to a power-law distribution alone \citep[e.g.][]{Mathis.1977, Davis.1990, Blum.1993, Guttler.2010}, the new model is also more accurate. In some of our simulations, a fragment distribution contains one large fragment and many small ones. The small ones can be accurately modelled with a power-law mass distribution. However, the mass of the largest fragment often does not match the power law. This is particularly true for grazing collisions between highly porous aggregates, where the filling factor of the largest remnant differs significantly from those of the other fragments.  Since this has a major impact on subsequent collisions, we treat the largest fragment and its filling factor separately. Furthermore, power-law distributions cannot map bouncing collisions. For these reasons, we distinguish the largest \emph{and} second largest fragment from the rest of the fragment distribution.

With the four given fragment populations and their defining properties, we present a closed model that is capable of modelling any collision outcome of pre-planetesimals with minimum complexity but with the accuracy necessary to model the dust aggregation in global coagulation models. In the next section, we show how the model by \citet{Guttler.2010} can be represented in our new model.

\subsection{Mapping sticking, bouncing, and fragmentation to the new model}
\label{sec:mapping}

\begin{table*}
\caption{Mapping sticking, bouncing, and fragmentation. In this table, we illustrate how the collision outcome types of Sec.~\ref{sec:collision-types} can be mapped to the four-population model using the characteristic quantities mass $m$ and filling factor $\phi$ as examples. The masses of the largest and second largest fragment, the power-law population, and the sub-resolution population are given by $\Mlst$, $\Msnd$, $\Mpwlw$, and $\Mdust$, respectively. We assume for the target mass $\mtar$ and projectile mass $\mproj$ before the collision $\mtar \ge \mproj$. The filling factors are represented by $\phitar$ for the target, $\phiproj$ for the projectile, $\philst$ for the largest fragment, and $\phisnd$ for the second largest fragment. We also give examples of the transition criteria between the types of Sec.~\ref{sec:collision-types}. A list of symbols can be found in the appendix and further explanations are given in the text.}             
\label{tab:mapping}      
\centering          
\begin{tabular*}{\hsize}{l@{\extracolsep{\fill}}c c c c c c }     
\hline\hline       
outcome type & $\Mlst$ & $\philst$ & $\Msnd$ & $\phisnd$ & $\Mpwlw$ & $\Mdust$ \\ 
\hline \\

hit \& stick (S1) & $\mtar + \mproj$ & $ \philst$ & 0 & 0 & 0 & 0 \\
              
sticking through surface effects (S2) & $\mtar + \mproj$ & $ \philst> \phitar$ & 0 & 0 & 0 & 0 \\

sticking by penetration (S3) & $\mtar + \mproj$ & $ \philst > \phitar$ & 0 & 0 & 0 & 0 \\

mass transfer (S4) & $\Mlst > \mtar$ & $ \philst > \phitar$ & $\Msnd < \mproj$ & $\phisnd > \phiproj$ & $\Mpwlw \ge 0$ & $\Mdust > 0$ \\

bouncing with compaction (B1) & $\Mlst = \mtar$ & $\philst > \phitar$ & $\Msnd = \mproj$ & $\phisnd > \phiproj$ & 0 & 0 \\

bouncing with mass transfer (B2) & $\Mlst \ne \mtar$ & $\philst > \phitar$ & $\Msnd \ne \mproj$ & $\phisnd > \phiproj$ & 0 & 0 \\

fragmentation (F1) & $\Mlst < \mtar$ & $\philst > \phitar$ & $\Msnd > 0$ & $\phisnd > \phiproj$ & $\Mpwlw > 0$ & $\Mdust > 0$ \\

erosion (F2) & $\Mlst \lesssim \mtar$ & $\philst > \phitar$ & $\Msnd \ge 0$ & $\phisnd > \phiproj$ & $\Mpwlw \ge 0$ & $\Mdust > 0$ \\

fragmentation with mass transfer (F3) & $\Mlst > \mtar$ & $\philst > \phitar$ & $\Msnd > 0$ & $\phisnd > \phiproj$ & $\Mpwlw > 0$ & $\Mdust > 0$ \\

sticking -- fragmentation transition & $\Mlst > \mtar \rightarrow \Mlst < \mtar$ & $\philst$ & $\Msnd \ge 0$ & $\phisnd$ & $\Mpwlw \ge 0$ & $\Mdust \ge 0$ \\

bouncing -- fragmentation transition & $\Mlst \le \mtar$ & $\philst$ & $\Msnd \ge 0$ & $\phisnd$ & $\Mpwlw = 0 \rightarrow \Mpwlw > 0$ & $\Mdust = 0 \rightarrow \Mdust > 0$ \\
\hline                  
\end{tabular*}
\end{table*}

The key idea behind mapping the sticking, bouncing, and fragmentation sub-types to the four-population model is to describe the collision outcome by the characteristic quantities of each of the four populations. If one of the populations does not exist, its characteristic quantity is simply set to zero. In principle, a number of quantities can be used to determine the population. To illustrate the use of this model, we select the characteristic quantities mass $m$ and filling factor $\phi$ as examples. Table \ref{tab:mapping} demonstrates the successful mapping of the \citet{Guttler.2010} types into our quantitative four-population model. The indices ``t'', ``p'', ``1'', ``2'', ``pw'', and ``sr'' denote target, projectile, largest, second largest, power-law, and sub-resolution quantities, respectively. For the illustration, we assume that the mass of the target $\mtar$ is greater than or equal to the projectile mass $\mproj$, i.e.\ $\mtar \ge \mproj$. We find a filling factor increase in all collisions if the initial filling factor is smaller than the maximum compaction at $\phi_{\rm max} = 0.58$ (see Fig.~\ref{fig:ff_lgst} for the largest fragment).

For pure sticking (S1 to S3 types) only one fragment in the final population exists, which is identified as the largest fragment. Consequently, this fragment contains the mass of the total system, i.e.\ $\Mlst = \mtar + \mproj$. Within the four-population approach, the S1 to S3 sub-types can be combined in terms of the mass and filling factor. In general, the filling factor of the target $\phitar$ is increased during the impact such that the filling factor of the largest fragment $\philst > \phitar$. For sticking with mass transfer (S4), a range of fragments exists. Since sticking is identified with growth of the largest fragment, it is $\Mlst > \mtar$. If there are only fragments below the resolution limit, then the mass of the second largest fragment $\Msnd$ and the mass of the power-law population $\Mpwlw$ vanish. Otherwise there are a range of fragments with non-zero masses. 

Bouncing is characterised by two fragments in the final fragment distribution. For bouncing with compaction (B1), the masses are unaltered, i.e., $\Mlst = \mtar$ and $\Msnd = \mproj$, but the filling factors are increased such that $\philst > \phitar$ and $\phisnd > \phiproj$. For the case of bouncing with mass transfer (B2), the filling factors are also increased but the masses are altered such that $\Mlst + \Msnd = \mtar + \mproj$ and $\Mlst \ne \mtar$ as well as $\Msnd \ne \mproj$. By definition, no fragments are produced in the bouncing events such that $\Mpwlw = \Mdust = 0$. For B2, the filling factors are also increased, i.e., $\philst > \phitar$ and $\phisnd > \phiproj$.

The fragmentation events (F1 to F2 types) are characterised by the mass of the largest fragment being lower than the highest mass before the collision, i.e.\ $\Mlst < \mtar$. The distinction between F1 and F2 in terms of the collision outcome is unclear. In both cases, a power-law as well as a sub-resolution population is generated, i.e.\ $\Mpwlw > 0$ and $\Mdust > 0$. It seems that for F2, the mass of the target is only reduced by a small amount such that $\Mlst \lesssim \mtar$ \citep{Guttler.2010}. In this reference, F3 is also characterised by the target (solid plate) gaining mass, i.e.\ $\Mlst > \mtar$. This is inconsistent with the idea that fragmentation is equivalent to $\Mlst < \mtar$. The mass of the second largest fragment has some value $\Msnd > 0$ and total masses of the power-law and sub-resolution populations are $\Mpwlw > 0$ and $\Mdust > 0$. 

The transition between sticking and fragmentation can consequently be characterised by the transition $\Mlst > \mtar \rightarrow \Mlst < \mtar$, since for the S4 type power-law and sub-resolution population are already present. In contrast, the transition from bouncing (which also includes the change of target and projectile masses by mass transfer in B2) can be defined by the appearance of a power-law and sub-resolution population, i.e.\ $\Mpwlw = 0 \rightarrow \Mpwlw > 0$ and $\Mdust = 0 \rightarrow \Mdust > 0$, together with a non-growing target $\Mlst \le \mtar$.

In this section, we have shown that the four-population model for collision outcomes is capable of encompassing all the sticking, bouncing, and fragmentation sub-types. Furthermore, we have demonstrated that the transition between these types can be modelled continuously by utilising the masses of each of the four populations.


\section{Applying the new model to simulation data}
\label{sec:application}

\begin{table}
\centering
\begin{tabular}{c c}
\hline\hline
Collision & Broad\\
Velocity [$\rm ms^{-1}$] & categorisation \\
\hline
0.1 & bouncing\\
0.2 & bouncing\\
1.0 & bouncing\\
1.25 & sticking\\
1.5 & sticking\\
1.75 & sticking\\
2.0 & sticking\\
2.2 & sticking\\
2.5 & sticking\\
4.0 & sticking\\
5.0 & sticking\\
6.0 & sticking\\
7.5 & sticking\\
10.5 & sticking\\
11.5 & fragmentation\\
12.5 & fragmentation\\
15.0 & fragmentation\\
17.5 & fragmentation\\
25.0 & fragmentation\\
27.5 & fragmentation\\
\hline
\end{tabular}
\caption{Collision velocities used in the simulations. The target and projectile are homogeneous and feature an intermediate porosity ($\phi = 0.35$). These simulations have been carried out to show how the proposed model can quantitatively demonstrate the results of pre-planetesimal collisions.  Each simulation is assigned a broad categorisation so that the transition regions between the categories can be investigated.}
\label{tab:application}
\end{table}

We apply the four-population model to a study of collisions of medium porosity pre-planetesimals with different velocities to demonstrate that the outcomes of pre-planetesimal collisions can be described quantitatively using the model.  We carry out 24 simulations of head-on collisions involving spherical dust aggregates with target and projectile radii of $\rt = \unit[10]{cm}$ and $\rp = \unit[6]{cm}$ modelled using 238,238 and 51,477 SPH particles, respectively. We choose the initial filling factor of the aggregates to be $\phi = 0.35$ resulting in target and projectile masses of 2.93 and \unit[0.63]{kg}, respectively. We vary the collision velocity between 0.1 and $\unit[27.5]{ms^{-1}}$.  Table~\ref{tab:application} summarises the simulations carried out.  We broadly categorise the simulations by eye into \emph{bouncing}, \emph{sticking}, or \emph{fragmentation} so that a comparison can be made between the results of our model and the categorisations used previously. We note that these broad categorisations are only employed to show where the boundaries between these regimes lie and are not used for the quantitative approach of our model.

As discussed in Sec.~\ref{sec:frag_pop}, the outcome of pre-planetesimal collisions can be quantitatively described by a number of parameters such as the final mass, size, energy, velocity, porosity, and rotation and each of these parameters can be used to describe the four populations.  In this preliminary, study we focus on the final mass, porosity, and energy of the different populations in our model.

\begin{figure}
            {\includegraphics[width=1.0\columnwidth]{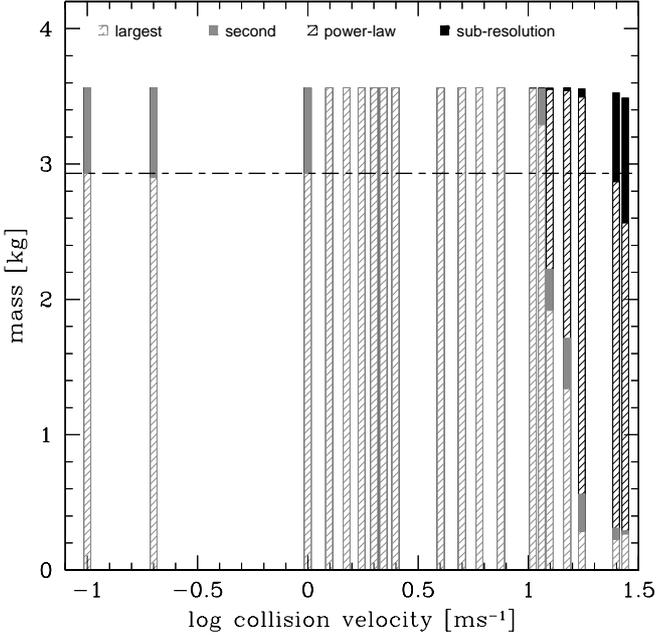}}
      \caption{Cumulative plot of the contributions to the total mass in the system by the largest (grey slashed), second largest (solid grey), power-law (black slashed), and sub-resolution (solid black) populations after the aggregate collisions at various velocities.  At low velocities, where bouncing occurs, the mass is wholly within the first and second largest fragments, while at intermediate velocities, where sticking occurs, the contribution to the mass is in the largest fragments.  At high velocities, the mass contribution from the power-law population becomes significant.}
         \label{fig:mass_distribution}
\end{figure}

Figure~\ref{fig:mass_distribution} shows how the mass distribution amongst the four populations varies with collision velocity.  At low collision velocities, the contact energy and thus the tensile strength is too low to hold the objects together.  This results in bouncing causing the overall mass distribution being similar to the initial distribution. The collision with $\unit[0.2]{\msec}$ shows evidence of mass transfer resulting in the largest fragment being smaller than the target. As the collision velocity increases, the energy is dissipated by plastic deformation resulting in sticking such that all the mass is stored in a single object.  At even higher collision velocities in which fragmentation is seen, the mass stored in the largest fragment is reduced but is increasingly present in the power-law population.  Figure~\ref{fig:mass_distribution} also shows how the mass of the largest fragment varies with collision velocity.  It can be seen that three distinct regions exist, which justify the broad classification of Table \ref{tab:application}. However, there do not appear to be any distinctions within these regions which justify an additional division into sticking, bouncing, and fragmentation sub-types. A sharp transition between the bouncing and sticking regions exists at $\unit[\sim 1]{\msec}$.  Above $\unit[10.5]{ms^{-1}}$, pure sticking no longer occurs and the mass of the largest fragment begins to decrease as the transition between sticking and fragmentation takes place. As a result of very high collision velocities, the mass contribution of the sub-resolution population dominates over the largest and second largest fragment which is an indicative of violent disruption. We note that the simulation with collision velocity $\unit[11.5]{ms^{-1}}$ still results in collision \emph{growth} but above this velocity the mass gain of the largest fragment is negative.

\begin{figure}
            {\includegraphics[width=1.0\columnwidth]{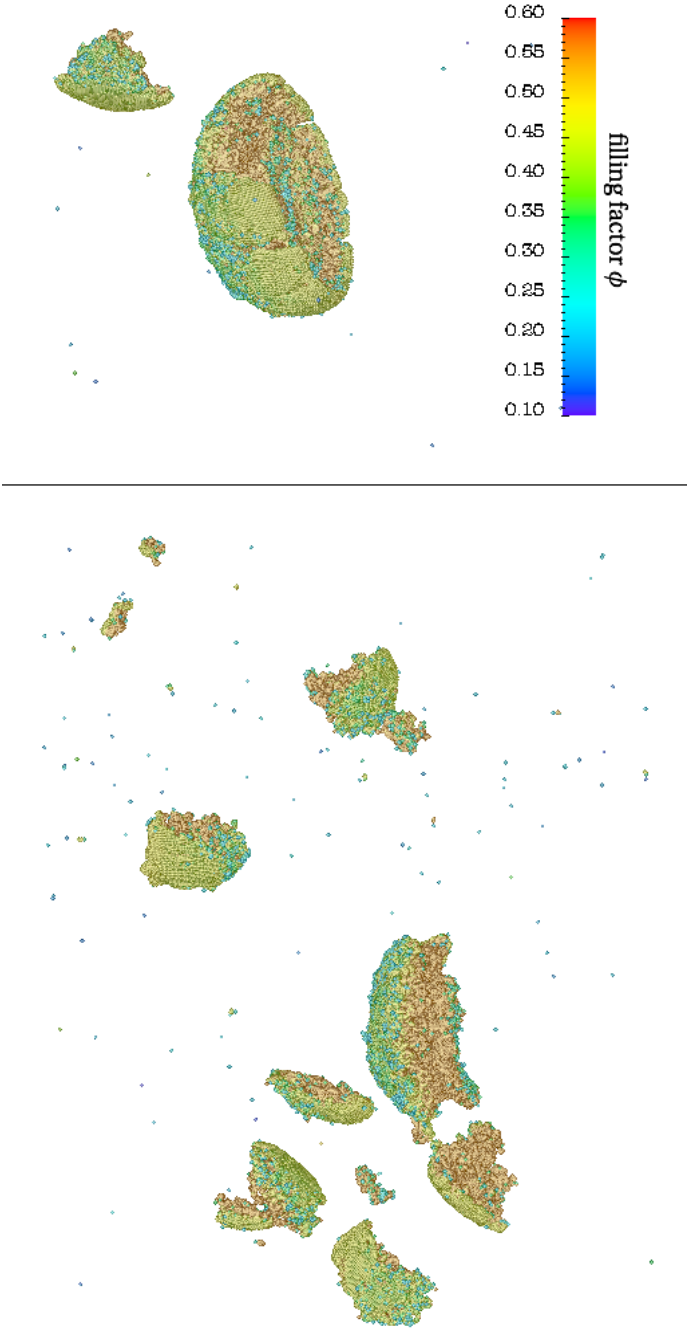}}
      \caption{Outcome of the dust collision simulation with $\unit[11.5]{ms^{-1}}$ (top) and $\unit[12.5]{\msec}$ (bottom) at a time, t=\unit[0.8]{s}.  The collision with $\unit[11.5]{ms^{-1}}$ results in the growth of the target but also a large chunk does break off such that pure sticking does not occur. The situation is adequately mapped with the appearance of a second largest fragment. For $\unit[12.5]{\msec}$ the target breaks apart and the power-law population becomes as significant as the largest fragment. The transition from sticking to fragmentation occurs in between the presented collision velocities.}
         \label{fig:image_11.5s}
\end{figure}

Figure~\ref{fig:mass_distribution} clearly shows that at velocities $\gtrsim \unit[11.5]{ms^{-1}}$, pure sticking no longer occurs and a transition into the \emph{fragmentation} regime begins.  For the simulation with a collision velocity of $\unit[11.5]{ms^{-1}}$ (see Fig.~\ref{fig:image_11.5s}, top, for an illustration of the resulting fragments), the mass of the largest fragment clearly increases but according to the model by \citet{Guttler.2010}, it is unclear whether this simulation would be classed as sticking with mass transfer (S4) or fragmentation with mass transfer (F3). Using our model, such an intermediate region can be described quantitatively.  At $\sim \unit[12.5]{ms^{-1}}$ (see Fig.~\ref{fig:image_11.5s}, bottom, for an illustration), the mass contribution from the power-law population becomes as significant as the contribution from the largest fragment.  In this region, the target fragments into a small number of large pieces.  We expect that the transition from the sticking to fragmentation regime, described in Section~\ref{sec:mapping}, occurs between 11.5 and $\unit[12.5]{ms^{-1}}$.

Using the model, it is clear that the mass of the largest fragment decreases sharply for collision velocities $\gtrsim \unit[11.5]{ms^{-1}}$, thus this is the region where fragmentation takes place. We expect that this fragmentation boundary will change depending on a number of other parameters such as impact parameter and mass ratio \citep{Blum.1993}, porosity, and the rotation of the dust aggregates.  Further investigation into the factors that determine the fragmentation boundary is crucial to ultimately understand under what conditions fragments may grow to planetesimal sizes.  It is important to note that to decrease the computational expense, the SPH particles that move out of a radius of \unit[20]{m} are removed from the simulation.  Therefore, the total mass at the end of the higher velocity simulations is smaller than the initial mass.  However, the total mass removed from the simulation is $\sim 2\%$ in the highest velocity simulation.


\begin{figure}
  {\includegraphics[width=1.0\columnwidth]{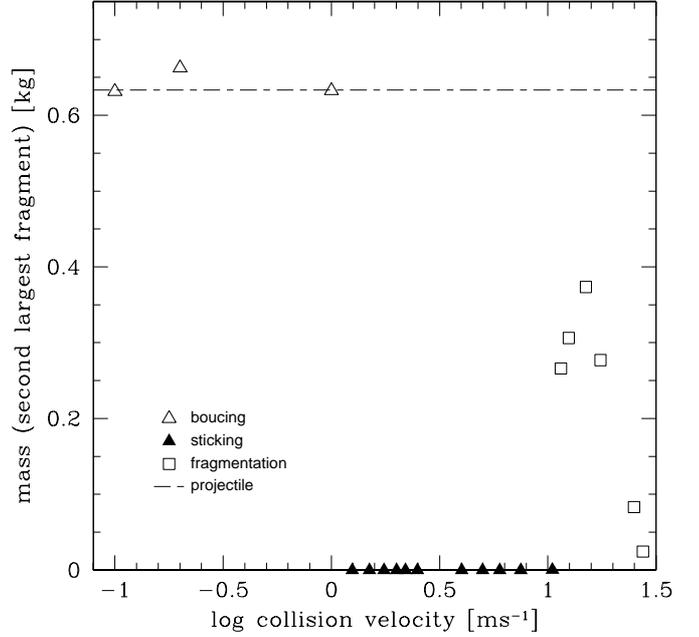}}
  \caption{Mass of the second largest fragments. The broad categorisation into \emph{bouncing} (open triangles), \emph{sticking} (closed triangles), and \emph{fragmentation} (open squares) can be identified by masses of the second largest fragments being approximately the projectile mass (dashed line), nearly vanishing, and smaller than the projectile mass, respectively. For the fragmenting collisions, the mass of the second largest fragment initially increases with collision velocity.  However, at higher collision velocities, the fragment mass decreases because of the collision nature of the simulations.}
  \label{fig:mass_second}
\end{figure}

Figure~\ref{fig:mass_second} shows how the mass of the second largest fragment varies. For low collision velocities, it is evident that \emph{bouncing} occurs and the projectile, whose mass is indicated by the dashed line in Fig.~\ref{fig:mass_second}, appears as the second largest fragment with unaltered or slightly increased mass in the final distribution. For higher collision velocities, the projectile \emph{sticks} to the target and no second largest fragment results, which is indicated by vanishing masses in this regime. As the velocity further increases, the resulting mass of the second largest fragment increases again but is smaller than the projectile mass, which indicates \emph{fragmentation}. The increase is followed by a sharp decrease in the mass of the largest fragment. These simulations, which are reasonably close to the sticking-fragmentation boundary not only cause the two aggregates to fragment, but also cause mass to be transferred between the target and projectile.  At higher velocities still, the impact is sufficiently violent for the second largest fragment to also decrease sufficiently.

\begin{figure}
  {\includegraphics[width=1.0\columnwidth]{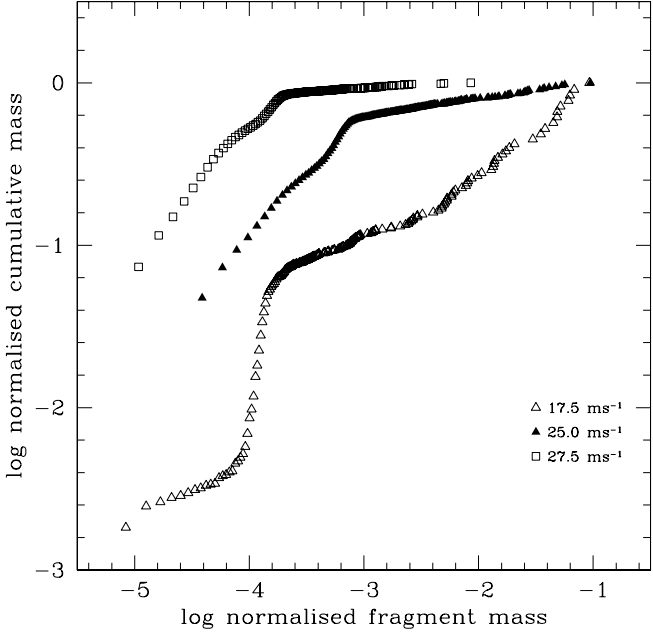}}
  \caption{Cumulative mass distribution (given by equation~\ref{eq:power-law}) against fragment mass of the power-law populations resulting from the simulations with 17.5 (open triangles), 25.0 (closed triangles), and $\unit[27.5]{ms^{-1}}$ (open squares).  Both axes are normalised by $\Mpwlw$. For high fragment masses, the gradient of the slope decreases with increasing velocity.  For low masses, it remains unclear whether the deviation from the power-law is a physical or numerical artefact.}
  \label{fig:pwlw_cum_mass}
\end{figure}

Figure~\ref{fig:mass_distribution} shows that as the collision velocity increases, the contribution to the mass from the power-law population also increases.  As an outcome of pre-planetesimal collisions, the cumulative mass distribution of the fragments is often described by a power-law
\begin{equation}
m_{\rm cum} (m_{\rm f}) = \sum_{m_{\rm f}<\mu_{\rm pw}} m_{\rm f} = \int_0^{m_{\rm f}} n(m) m \, \rmd m= \left( \frac{m_{\rm f}}{\mu_{\rm pw}} \right)^\kappa \; ,
\label{eq:power-law}
\end{equation}
where $n(m) \, \rmd m$ is the number of fragments within the mass range $[m, m + \rmd m]$, $m_{\rm f}$ is the fragment mass, and $\mu_{\rm pw}$ is the most massive member of the power-law population. The quantities $m$, $m_{\rm f}$, and $\mu_{\rm pw}$ are normalised by the total mass of the power-law population $\Mpwlw$ and $\kappa$ is the power-law index. 

Figure~\ref{fig:pwlw_cum_mass} shows the cumulative mass distribution $m_{\rm cum} (m_{\rm f})$ against the fragment mass normalised by $\Mpwlw$. For higher fragment masses $m_{\rm f}$, a power-law fit can be obtained that describes the mass distribution of the power-law population.  As the velocity increases, the slope of the power-law decreases.  This is because at higher collision velocities, the destruction of the dust aggregates is more violent and a larger number of smaller fragments results.  At low fragment masses, it is currently unclear whether the deviation from the power-law distribution is physical, or whether it is a result of a low number of SPH particles ($\lesssim 100$) per fragment.  We note that this deviation is also seen in the experimental results of \citet{Guttler.2010}.

\begin{figure}
            {\includegraphics[width=1.0\columnwidth]{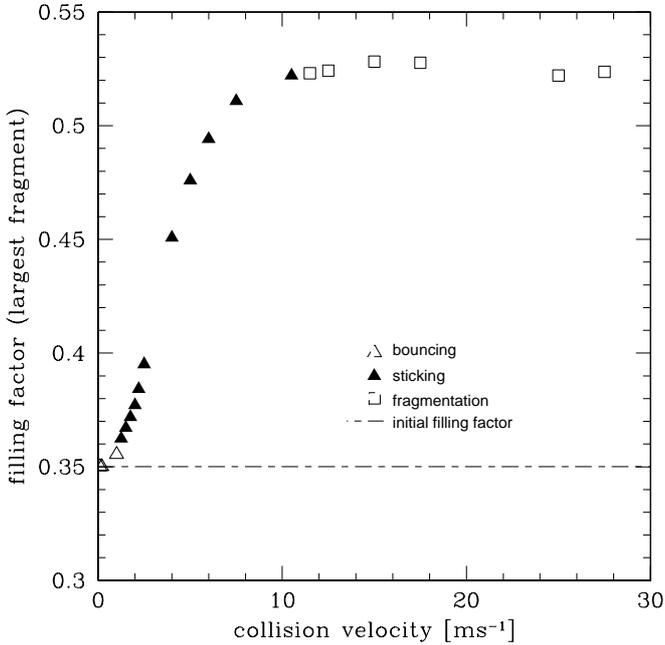}}
      \caption{Final average filling factor of the largest fragment in each of the simulations.  The simulations have been broadly categorised into bouncing (open triangles), sticking (solid triangles) and fragmentation (open squares).  The filling factor varies smoothly as the collision velocity is increased.}
         \label{fig:ff_lgst}
\end{figure}

We now consider how much compaction takes place in each of the simulations by considering the resulting filling factors of the various populations.  Figure~\ref{fig:ff_lgst} shows the average filling factor of the largest fragment, $\phi_1$, compared to the initial filling factor.  As the velocity increases, the compaction also increases, causing the filling factor to increase with velocity.  This forms a smooth curve between the initial filling factor of 0.35 and the maximum filling factor of $\sim 0.58$ for $\rm SiO_2$. The curve resembles the Fermi-Dirac shape of the compressive strength relation (Eq.~\ref{eq:compressive-strength}), which directly links the dynamic pressure to the obtained filling factor. We note that the appearance of fragmenting events coincides with the largest fragment reaching a filling factor close to the maximum filling factor. This suggests that fragmentation sets in when the material is maximally compressed. Furthermore, we stress that the transition of the filling factor is smooth right from the low velocity (bouncing) collisions, through to the medium velocity (sticking) collisions to the high velocity (fragmentation) collisions.  For simulations with higher initial filling factors, we expect there to be a direct transition between the bouncing and fragmentation regimes, as discussed in Section~\ref{sec:mapping}.

The transition is particularly important because for planet formation, we are primarily concerned with the evolution of the largest fragment.  This demonstrates that our model can quantitatively capture the results of the collisions over a wide region of velocity parameter space.  Our preliminary results for the second largest fragment suggest that only a small amount of compaction takes place for the second largest fragment in low velocity \emph{bouncing} simulations such that the final filling factor is very close to the initial filling factor, while for high velocity \emph{fragmentation} simulations, a large amount of compaction takes place resulting in filling factors close to the maximum value of 0.58.  Additional investigations will be carried out in a future study.

\begin{figure}
            {\includegraphics[width=1.0\columnwidth]{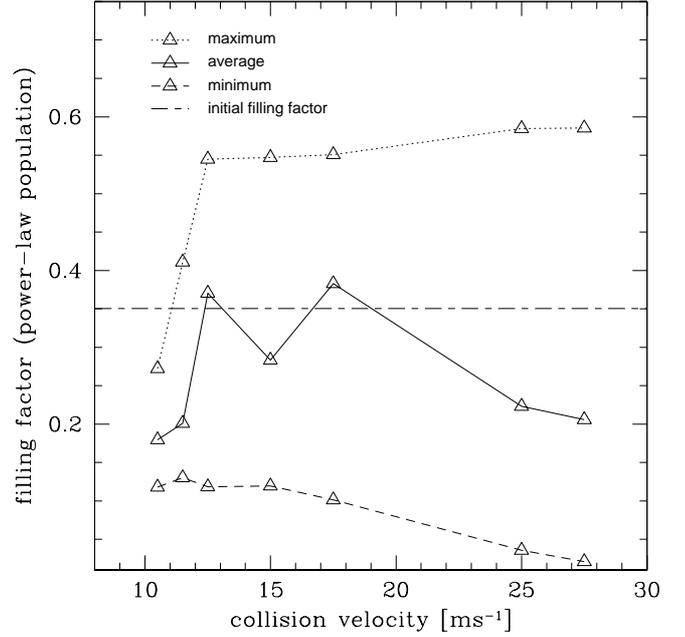}}
      \caption{Final average filling factor (black line) of the power-law populations in each of the simulations.  We also plot the maximum (upper grey line) and minimum (lower grey line) filling factors for each simulation, as well as the initial filling factor (dashed line).  As the collision velocity increases, the range of filling factors for any one simulation increases.}
         \label{fig:ff_pwlw}
\end{figure}

\begin{figure}
            {\includegraphics[width=1.0\columnwidth]{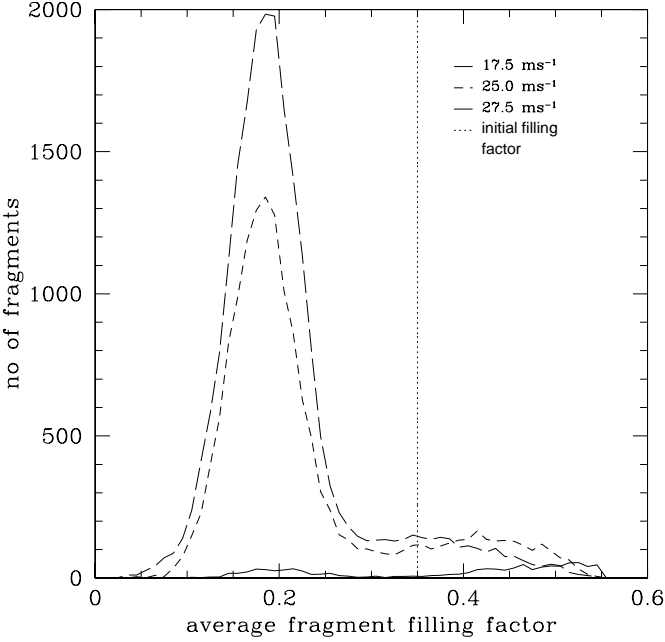}}
      \caption{Filling factor distribution for the simulations with 17.5 (solid line), 25.0 (short dashed line) and $\unit[27.5]{ms^{-1}}$ (long dashed line). The size of a filling factor bin is 0.01. As the collision velocity increases, the number of fragments increases.  For higher collision velocities, the majority of fragments have lower filling factors than the original value (dotted line).}
         \label{fig:ff_dis_pwlw}
\end{figure}

Figure~\ref{fig:ff_pwlw} shows the average, maximum, and minimum filling factors of the power-law population.  At higher velocities when more violent fragmentation occurs, the spread in the filling factor is greater than at lower velocities.  This suggests that at lower velocities, fragments may be chipped off without too much change to their compaction, whereas in the more violent cases with higher collision velocities, the fragments may become compacted before being chipped off or rupture because the plastic flow occurs where parts of the dust are stretched before being ripped off.  Figure~\ref{fig:ff_dis_pwlw} shows the filling factor distribution of the power-law populations resulting from the simulations with 17.5, 25.0, and $\unit[27.5]{ms^{-1}}$.  As the collision velocity increases, the number of fragments increases.  In particular, most of the particles have filling factors smaller than the initial value of 0.35.  This suggests that during the collision, fragments are ripped off instead of being compressed before breaking apart.

\begin{figure}
            {\includegraphics[width=1.0\columnwidth]{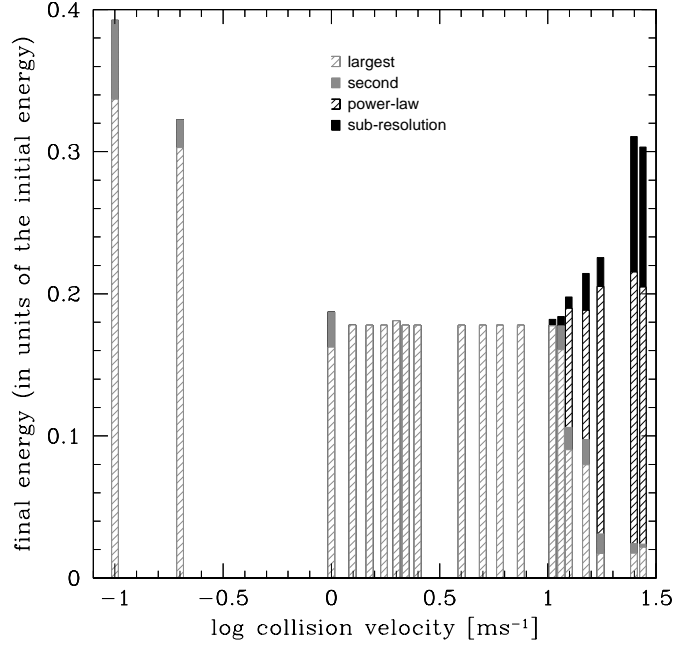}}
      \caption{Cumulative plot of the contributions to the total energy in the system by the largest (grey slashed), second largest (solid grey), power-law (black slashed), and sub-resolution (solid black) populations after the aggregate collisions at various velocities.  At low velocities, the energy is primarily from the largest fragment, whereas at higher collision velocities, where more fragments form, the contribution from the power-law population becomes more significant.}
         \label{fig:energy_distribution}
\end{figure}

\begin{figure}
            {\includegraphics[width=1.0\columnwidth]{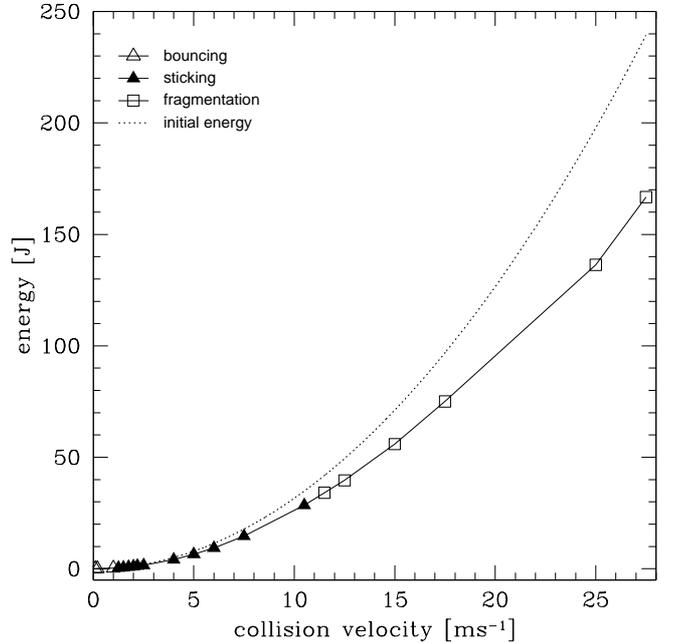}}
      \caption{Initial (dotted line) and dissipated (solid line) energies against the collision velocity.  The amount of energy dissipated changes smoothly with collision velocity, regardless of whether the simulations involves bouncing (open triangles), sticking (closed triangles), or fragmentation (open triangles).}
         \label{fig:energy_init_diss}
\end{figure}

Finally, we explore the contributions to the total energy (normalised by the initial energies) from the different populations.  The total energy is the sum of the energy stored in translation, rotation, and vibration. Figure~\ref{fig:energy_distribution} shows that at low velocities the major energy contribution comes from the largest fragment, while at higher velocities the contribution from the power-law population becomes somewhat equally significant (at $\sim \unit[12.5]{ms^{-1}}$) and even more so at even higher velocities.  At low velocities, the energy is stored in elastic deformation and released into kinetic energy again so that a smaller fraction of the initial energy is dissipated.  As the velocity increases plastic deformation occurs causing energy dissipation.  At even higher velocities, though energy is still dissipated (since bonds are broken in the fragmentation process), the dissipated energy as a fraction of the initial energy due to bonds being broken is not as large as the energy dissipation due to plastic deformation.  For very high impact velocities, a considerable amount of energy is stored in the sub-resolution population. Figure~\ref{fig:energy_init_diss} shows the absolute energy dissipated at each collision velocity as well as the initial energy.  The total amount of energy dissipated increases smoothly with increasing velocity, irrespective of the \emph{bouncing}, \emph{sticking}, or \emph{fragmentation} regime.  We note that in these examples, the final rotational energies are not very significant because the collisions are head-on.  The difference between the rotational and translational energies will become more important in simulations with non-zero impact parameters.

We have presented here the results of a small number of simulations, which show that transitions between the \emph{bouncing}, \emph{sticking}, and \emph{fragmentation} regions occur and that the results of pre-planetesimal collisions can be described quantitatively. Our study could be extended to explore a vast parameter space of inputs.  Our model has been shown not only to encompass previous qualitative models (Sec.~\ref{sec:new-model}) but also provide an alternative model that focuses on the quantitative outcome of pre-planetesimal collisions.  Furthermore, the model can provide an accurate input into global models, which may then carry out detailed calculations of pre-planetesimal collisions.  In addition, this model provides a powerful aid to determine the region of parameter space that allows growth of pre-planetesimals to occur.


\section{Discussion and outlook}
\label{sec:discussion}

We have demonstrated that our SPH code, extended and calibrated for the simulation of porous $\rm SiO_2$ dust, is capable of reproducing all sticking, bouncing, and fragmentation types that appear in collision experiments with macroscopic porous dust aggregates. In addition to the quantitatively correct simulation of laboratory benchmark experiments \citep{Guttler.2009, Geretshauser.2010}, this consolidates the validity of the applied porosity model and shows its readiness for the application in the field of investigating pre-planetesimal collisions. Since the continuum approach of the SPH method does not place an upper bound on the aggregate size, collision data of a parameter space that have been inaccessible so far in laboratory experiments can now be provided for further use in global coagulation models.

For this transfer, a suitable mapping of collision data has to be chosen, which is accurate enough to quantitatively capture the most important features of any combination of sticking, bouncing, and fragmentation and simple enough to be implementable in global coagulation models. We have attempted to map our simulation data to the categorisation by \citet{Guttler.2010}, which was the most elaborate collision model previously available. On the one hand, the distinction between four sticking, two bouncing, and three fragmentation types introduces a considerable amount of complexity caused by distinguishing between a mixture of qualitative and quantitative attributes and by adhering to a distinct classification of these events. On the other hand, we have found a collision outcome that could not clearly be attributed to one of the proposed categories.

Because of this ambiguity, we have proposed a new model, which is based on quantitative aspects. For this purpose, we have divided the set of fragments of a collision into four populations: the largest and second largest fragment are described by distinct values for the characteristic quantities of mass, filling factor, and kinetic energy to name only a few. The power-law population is described by distributions and the sub-resolution population by average values of the characteristic quantities. The largest fragment indicates growth or erosion, the second largest fragment accounts for bouncing, the power-law population quantitatively describes the amount of fragmentation, and the sub-resolution population gives an upper limit for smaller fragments, which are not captured because of insufficient resolution. Since the SPH code is not restricted to small aggregate sizes, the importance of the sub-resolution population becomes significant for aggregate collisions between objects of approximately metre size and more. For growth models that rely on the sweeping up of small particles \citep{Teiser.2009}, the sub-resolution population also plays an important role. We have demonstrated that this model in general encompasses the model proposed by \citet{Guttler.2010} but is also capable of capturing intermediate events.

Finally we have applied the new model to map data for head-on collisions of aggregates with intermediate ($\phi = 0.35$) porosity and varying impact velocity. We have shown that the broad sticking, bouncing, and fragmentation categorisation can still be found in the four-population model. But in addition, we have also shown that continuous transitions in the variation in the mass of the largest fragment, the filling factor, and the final kinetic energy of the fragments with collision velocity exhibit a more quantitative description. The ability to capture these transitions justifies the design of the four-population model and demonstrates its descriptive power.

Despite its narrow parameter range, we can draw the following conclusions from the velocity study of Sec.~\ref{sec:application}. For the transition between bouncing and sticking, we find a threshold velocity of $\unit[\sim 1]{\msec}$ and that the transition from sticking to fragmentation lies between 11.5 and $\unit[12.5]{\msec}$. Since our initial filling factor $\phi = 0.35$ is close to the critical filling factor 0.4, which separates porous from compact aggregates in \citet{Guttler.2010}, our findings cannot be directly compared to their results. For collisions between equally sized highly porous and very compact aggregates, they found a direct transition from bouncing to fragmentation at $\unit[\sim 1]{\msec}$. However, for collisions between a compact target and a porous projectile of equal size they find a bouncing-sticking transition at $\unit[1]{\msec}$ and a transition from sticking to no mass gain at $\unit[9.4]{\msec}$. These thresholds resemble our findings for intermediate porosity very well. Our results for fragmenting collisions indicate that the power-law index of the fragment mass distribution is velocity dependent. This is supported by the collection of laboratory fragmentation data \citep{Mathis.1977, Davis.1990, Blum.1993, Guttler.2010}. Furthermore, it is physically reasonable that in more violent collisions the fraction of smaller fragments increases. This suggests a velocity-dependent power-law index.

We note that the results presented in Sec.~\ref{sec:application} are valid for collisions between homogeneous aggregates of intermediate porosity. Increasing inhomogeneity might affect the presented threshold velocities. Furthermore, rotation of the largest and second largest fragments might cause them to fall apart beyond the simulated time, an outcome that eventually affects their final size. The results presented here, in particular for fragmenting collisions, are based on one simulation for each collision velocity. Although we do not expect a large variation in outcome because of the symmetry given by a head-on collision of two spheres, a profound statistical investigation has to be carried out. It remains to be investigated whether the deviation of the power-law mass distribution from a power-law for low fragment masses is a numerical or physical effect. Despite these drawbacks, we have been able to demonstrate the applicability and functionality of the four-population model by means of our simulation results.

Despite increased experimental efforts, only small spots of the required parameter space of pre-planetesimal collisions are actually covered by empirical data. Vast regions of these maps remain \emph{terra incognita} and collision data for pre-planetesimal sizes larger than centimetre are missing.

With both a code calibrated for the simulation of pre-planetesimals \citep{Geretshauser.2010} and the four-population model as an adequate mapping model, we have established a basis for profoundly investigating all aspects of pre-planetesimal collisions and transferring acquired data to global dust coagulation models. In future studies, these tools can be utilised to generate a catalogue of pre-planetesimal collisions. The four-population model will be applied to investigate how the collision behaviour depends on important parameters such as aggregate porosity and inhomogeneity, mass ratio of the collision partners, impact velocity, impact parameter, and rotation. Furthermore, we will assess the statistics of our fragmentation results.

\begin{acknowledgements}
R.J.G. and R.S. wish to thank C.\ G\"uttler, M.\ Krause, and J.\ Blum for the intensive and fruitful collaboration and W.\ Benz and his group for many illuminating discussions. We thank our anonymous referee for the thorough revision and constructive comments, which helped to improve this paper. The SPH simulations were performed on the university cluster of the computing centre of T\"ubingen, the bwGriD clusters in Karlsruhe, Stuttgart, and T\"ubingen. Computing time was also provided by the High Performance Computing Centre Stuttgart (HLRS) on the national supercomputer NEC Nehalem Cluster under project grant SPH-PPC/12848. This project was funded by the Deutsche Forschungsgemeinschaft within the Forschergruppe ``The Formation of Planets: The Critical First Growth Phase" under grant Kl 650/8-1.
 \end{acknowledgements}

\begin{appendix}

\section{List of symbols}

\begin{table}
\caption{List of symbols}             
\label{tab:symbol-list}      
\centering                          
\begin{tabular*}{\hsize}{l@{\extracolsep{\fill}}l}        
\hline\hline                 
symbol & explanation \\    
\hline                        
\\
$b$				& impact parameter \\
$E_{\rm rot,t}$		& rotational energy of the target \\
$E_{\rm rot,p}$		& rotational energy of the projectile \\
$J_2$			& second irreducible invariant\\
$K(\phi)$			& bulk modulus \\
$K_0$			& bulk modulus for the RBD dust sample \\
$m$				& mass \\
$m_{\rm cum}$		& cumulated mass \\
$m_{\rm f}$		& fragment mass \\
$\mproj$			& projectile mass \\
$\Mpwlw$			& mass of the total power-law fragment population \\
$\Mdust$			& mass of the total sub-resolution population \\
$\mtar$			& target mass \\
$\Mlst$			& mass of the largest fragment \\
$\Msnd$			& mass of the second largest fragment \\
$p$				& hydrostatic pressure \\
$\pmean$			& mean pressure of $\Sigma$ \\
$\Sab$			& deviatoric stress tensor \\
$T(\phi)$				& tensile strength \\
$\va$			& velocity vector \\
$v_0$			& relative collision velocity \\
$\xa$			& position vector \\
$Y(\phi)$			& shear strength \\
$\Delta$			& slope of $\Sigma$ \\
$\kappa$			& index of the power-law mass distribution \\
$\mu (\phi)$		& shear modulus \\
$\mu_{\rm pw}$		& most massive member of the power-law population \\
$\rho$			& density \\
$\stab$			& stress tensor \\
$\phi$			& filling factor \\
$\phi_c^+$		& threshold filling factor from elastic to plastic compression \\
$\phi_c^-$			& threshold filling factor from elastic to plastic tension \\
$\phi_{\rm max}$	& maximum filling factor of $\Sigma$ \\
$\phi_{\rm min}$	& minimum filling factor of $\Sigma$ \\
$\phiproj$			& filling factor of the projectile \\
$\phi_{\rm RBD}$	& filling factor of an uncompressed RDB dust sample \\
$\phitar$			& filling factor of the target \\
$\philst$			& filling factor of the largest fragment \\
$\phisnd$			& filling factor of the second largest fragment \\
$\Sigma(\phi)$		& compressive strength \\

\hline                                   
\end{tabular*}
\end{table}

\end{appendix}

\bibliographystyle{aa}
\bibliography{literature}

\end{document}